# The tunnelling spectra of quasi-free-standing graphene monolayer

Si-Yu Li, Ke-Ke Bai, Wei-Jie Zuo, Yi-Wen Liu, Zhong-Qiu Fu, Wen-Xiao Wang, Yu Zhang, Long-Jing Yin, Jia-Bin Qiao, Lin He*

Department of Physics, the Center for Advanced Quantum Studies, Beijing Normal University, Beijing, 100875, People's Republic of China

*Correspondence and requests for materials should be addressed to L.H. (e-mail: helin@bnu.edu.cn).

**With considering the great success of scanning tunnelling microscopy (STM) studies of graphene in the past few years, it is quite surprising to notice that there is still a fundamental contradiction about the reported tunnelling spectra of quasi-free-standing graphene monolayer. Many groups observed "V-shape" spectra with linearly vanishing density-of-state (DOS) at the Dirac point, whereas, the others reported spectra with a gap of ±60 meV pinned to the Fermi level in the quasi-free-standing graphene monolayer. Here we systematically studied the two contradicted tunnelling spectra of the quasi-free-standing graphene monolayer on several different substrates and provided a consistent interpretation about the result. The gap in the spectra arises from the out-of-plane phonons in graphene, which mix the Dirac electrons at the Brillouin zone corners with the nearly free-electron states at the zone center. Our experiment indicated that interactions with substrates could effectively suppress effects of the out-of-plane phonons in graphene and enable us to detect only the DOS of the Dirac electrons in the spectra. We also show that it is possible to switch on and off the out-of-plane phonons of graphene at the nanoscale, i.e., the tunnelling spectra show switching between the two distinct features, through voltage pulses applied to the STM tip.**

In the past few years, scanning tunnelling microscopy (STM) has been widely used to study the electronic properties of graphene[1-32]. Numerous exciting physical phenomena, such as strain-induced pseudo-Landau quantization[9,10], emergence of topological edge states in AB-BA domain wall[16], quasi-bound states in graphene quantum dots[22-27], and unconventional magnetism[28-32], have been observed in graphene systems via using STM. Even though the STM studies of graphene have achieved great success, there is still a fundamental contradiction about the observed tunnelling spectra of quasi-free-standing graphene monolayer. In most of the STM measurements, the reported spectra are "V-shape" and exhibit vanishing density-of-state (DOS) at the Dirac point[1-4,14], as expected for Dirac electrons in pristine graphene monolayer. However, an unexpected gap of ±60 meV pinned to the Fermi level is also observed in the tunnelling spectra of quasi-free-standing graphene monolayer in many experiments[5-9,15,24]. The two results are, obviously, contradicted and a consistent understanding of the tunnelling spectra of the quasi-free-standing graphene monolayer is still lacking after more than ten years STM studies of graphene. In this work, we systematically studied the two distinct tunnelling spectra of the quasi-free-standing graphene monolayer on several different substrates and showed that they can be understood consistently. We also demonstrated that it is possible to switch the tunnelling spectra between the two distinct features of graphene at the nanoscale through voltage pulses applied to the STM tip.

The gapped tunnelling spectra of graphene monolayer arise from band mixing between the linear dispersion quasiparticles (massless Dirac fermions) from Brillouin zone corners and higher energy excitations (nearly free-electron states) at the zone center[5,33]. The $K$ and $K'$ out-of-plane phonons of graphene play a key role in the band mixing, and decoupling of the graphene monolayer from the substrates (or quite weak interaction between graphene and the substrates) is crucial for the observation of the gapped tunnelling spectra[5-9,15,24,33]. In our experiment, we observed the gapped spectra at the nanoscale suspended graphene regions on different substrates, such as Rh foils, Pd foils, and single crystal $SrTiO_3$ substrates (see Methods and supplemental Fig. 1 for the growth of the samples). Figure 1 shows a representative STM and scanning tunnelling spectroscopy (STS) study of a suspended graphene region on a Rh foil. In Fig. 1a, we show a STM image of the nanoscale suspended graphene region on the Rh foil. The spectra recorded in the suspended graphene region, as shown in Fig. 1b, exhibit a ~ 116 mV gap-like feature centred at the Fermi energy, which is similar as that observed in previous STM studies[5-9,15,24]. The value of the gap is almost independent of the sample bias and the tunneling current during the STS measurements, as shown in Fig. 1c (we can also deduce the gap from the corresponding $d^2I/dV^2$ spectra shown in supplemental Fig. 2). Such a gap-like feature is attributed to the phonon-mediated tunneling and the steps in the differential tunneling conductance reflect the excitation energy of the phonons, as demonstrated explicitly in previous studies[5-9,15,24,33]. The measured excitation energy of the phonons, ~ 58 meV, in our STS measurements agrees quite well with the energy of the $K$ and $K'$ out-of-plane phonons of graphene, 56-58 meV, obtained from inelastic electron tunneling spectroscopy (IETS) measurements and density functional theory (DFT) calculations[8]. Similar results have also been observed in suspended graphene regions on Pd foils and SrTiO$_3$ substrate in our experiments (see supplemental Fig. 3 and Fig. 4), indicating that the gapped tunnelling spectrum is really a universal feature of the quasi-free-standing graphene monolayer.

Even though the gapped tunnelling spectrum is believed to mainly reflect the DOS of the nearly free-electron states, the Dirac electrons still have their contribution in the obtained differential tunneling conductance[5,33]. To explicitly confirm this, we measured evolution of the gapped spectra of the graphene monolayer with increasing magnetic fields for the first time. In the presence of perpendicular magnetic fields, it is expected to observe the Landau quantization of the Dirac fermions in graphene[1-4,14]. Figure 1d and 1e show a series of tunneling spectra for various magnetic fields recorded on the suspended graphene region on the Rh foil. Obviously, the Landau levels (LLs) developed as the magnetic fields are increased and four LL peaks $n$ = 1, 0, -1, -2 are well resolved in the magnetic field of 8 T. For massless Dirac fermions in graphene monolayer, the observed LL energies $E_n$ should depend on the square-root of both level index $n$ and magnetic field $B$:

$$E_n = \text{sgn}(n)\sqrt{2e\hbar v_F^2 |n| B} + E_0, \qquad n = ...-2,\ -1,\ 0,\ 1,\ 2... \qquad (1)$$

Here $E_0$ is the energy of Dirac point, $e$ is the electron charge, $\hbar$ is the Planck's constant, $v_F$ is the Fermi velocity, and $n > 0$ corresponds to empty-state (holes) and $n < 0$ to filled-state (electrons)[1-4,14]. The sequence of the observed Landau levels, as shown in Fig. 1f can be described quite well by Eq. (1). The linear fit of the experimental data to Eq. (1) allow us to obtain the Fermi velocities of electrons $v_F^e$ = (1.33 ±0.03) × 10$^6$ m/s and holes $v_F^h$ = (1.00 ±0.03) × 10$^6$ m/s separately. Similar large electron-hole asymmetry has been observed previously in strained graphene monolayer and is mainly attributed to the enhanced next-nearest-neighbor hopping due to the lattice deformation and out-of-plane curvature in the studied samples[31,34,35].

Based on the high-field spectra of the quasi-free-standing graphene monolayer (Fig. 1d-1e), two important results can be obtained. First, the observation of the Landau quantization of massless Dirac fermions demonstrated explicitly that the low-energy DOS of the quasi-free-standing graphene monolayer is still dominated by the Dirac electron's DOS even when the tunneling spectra (i.e., the tunneling DOS) exhibit the gap-like feature. Second, the signals of the LLs outside the gap are much stronger than that inside the gap, as shown in Fig. 1d and 1e, indicating that the tunneling DOS of the Dirac electrons outside the gap is also enhanced by the $K$ and $K'$ out-of-plane phonons. At present, it was believed that only the tunneling DOS outside the gap mainly reflects the contribution from the nearly free-electron states[5,33]. Our experimental result, obviously, is beyond the current understanding about the gapped tunneling spectrum of the quasi-free-standing graphene monolayer. Further theoretical analysis by taking into account effects of the $K$ and $K'$ out-of-plane phonons on the Dirac electrons is needed to fully understand the gapped tunneling spectrum.

Because of the $K$ and $K'$ out-of-plane phonons, the STM spectra not only reflect the contribution of the Dirac fermions in graphene[5,33]. Such an effect, to some extents, hampers the studies of the Dirac fermions in graphene by using STM. To detect only the Dirac fermions in the tunneling spectrum, we have to suppress the effects of the $K$ and $K'$ out-of-plane phonons of graphene. Our experiments demonstrated that interaction with substrates could efficiently suppress the effects of the out-of-plane phonons and enable us to achieve this goal. Figure 2 shows a success example of such a STM study. The studied sample is graphene multilayer on Rh foil (see method) and there is usually rotational misalignment between the adjacent graphene layers, which results in moiré patterns in STM image (Fig. 2a). Because of the large rotation angle $\theta$ ~ 16.3°, the topmost graphene sheet can electronically decouple from the underlying graphene systems[1-4,14,31]. Figure 2b shows tunneling spectra of the graphene sheet in the presence of different magnetic fields. In zero magnetic field, the spectrum exhibit the "V" shape characteristic for massless Dirac fermions in graphene monolayer rather than the gap-like feature. With increasing magnetic fields, we observed high-quality well-defined Landau levels of the massless Dirac fermions, as shown in Fig. 2b and 2c, indicating that the Dirac fermions dominate the STM spectra in such a case. Similar behavior has also been observed in decoupled graphene monolayer on graphite and graphene multilayer grown on other substrates, such as Ni foils, Pd foils, and SiC in our experiment (see supplemental Fig. 5-Fig. 8). Therefore, our results demonstrated that even the weak van der Waals interaction between the topmost graphene monolayer and the underlaying graphene (or graphite) could effectively suppress the effects of out-of-plane phonons and allow us to detect only the Dirac fermions in the STM studies.

Besides the graphite or graphene multilayer, the interaction between graphene and other substrates, such as metallic surface, could also effectively suppress the effects of the out-of-plane phonons. Figure 3 summarizes STM measurements of a graphene monolayer on a Cu(111) surface. It is well-known that the bands of Dirac fermions is fully intact with linear dispersion preserved for the graphene monolayer on the Cu(111) surface because of their weak interaction[36]. The moiré patterns with the period of 2.5 nm, as shown in Fig. 3a, arise from stacking misorientation between the graphene sheet and the Cu(111) surface. Figure 3b shows representative STS spectra of the graphene monolayer on the Cu(111) surface. Obviously, the spectra do not exhibit the gap-like feature and only show DOS for the Dirac fermions. A local minimum of the tunneling conductance, marked by the dashed line in Fig. 3b, is attributed to the Dirac point $E_D$ of graphene, which is in good agreement with that reported previously for the Dirac point of graphene on the Cu(111) surface[24,25]. Even though we can detect only the Dirac fermions' DOS for graphene monolayer on the Cu(111) surface, we cannot observe Landau quantization of the Dirac fermions even in the presence of perpendicular magnetic field of 8 T (Fig. 3b). Such a result is reasonable because that the electrons can tunnel between graphene and the Cu substrate, i.e., the graphene sheet is not electronically decoupled from the Cu(111) surface[24,25]. Therefore, due to the existence of the conducting substrate (the Cu surface in our experiment), the electrons in graphene on longer behave as an ideal two-dimensional free electron gas and consequently we cannot observe Landau quantization in the graphene even in the presence of large perpendicular magnetic fields.

Based on the above experimental results, we can conclude that both the "V-shaped" spectra and the gapped spectra are "correct" tunneling spectra of the quasi-free-standing graphene monolayer. However, the "V-shaped" spectrum only exhibits the contribution of the low-energy Dirac fermions, whereas the gapped spectrum is contributed by both the high-energy nearly free-electron states and the low-energy Dirac fermions. The two distinct spectra reflect the different tunneling DOS in the absence or presence of the *K* and *K′* out-of-plane phonons of graphene. Our results shown in Figures 1-3 demonstrated that we can observe the two distinct tunneling spectra of the quasi-free-standing graphene monolayer because of the different interactions between graphene and the supporting substrate. It implies that the interaction between graphene and the substrate can be used as an effective "switch" to turn on/off the effects of the *K* and *K′* out-of-plane phonons.

To further explore the effects of the interaction between graphene and the substrate on the tunneling spectra, we further carried out following controlled experiments, as shown in Fig. 4, in the graphene monolayer grown on a Rh foil. Our experiment indicated that we can directly tune the two distinct spectra of the graphene monolayer, i.e., we can manipulate the interaction between graphene and the substrate, with the STM tip. Figure 4a presents a STM image of a suspended graphene region on a Rh foil, which exhibits the gapped tunneling spectrum (Fig. 4c). To change the tunneling spectrum, the STM tip was positioned about 1 nm above the suspended graphene area and a bias of 4 V was applied for 0.2 second. After applying this voltage pulse, a STM image was measured over the same region, as shown in Fig. 4b. Obviously, the morphology of the suspended graphene area changes a lot after the voltage pulse. More importantly, STS spectrum acquired on part of the graphene area, as shown in Fig. 4c, no longer show the characteristic features of the gapped tunneling spectrum, indicating that the effect of the *K* and *K′* out-of-plane phonons is suppressed in part of the studied graphene area (the area between two dashed lines in Fig. 4b). Such a result is attributed to the enhanced interaction between the substrate and part of the studied graphene area after applying the voltage pulse. Figure 4d shows a STS map recorded in the same region of Fig. 4b within the energy of the out-of-plane phonons (the gap of the spectra). The region with high contrast in the STS map is the area where the effect of the *K* and *K′* out-of-plane phonons is effectively suppressed after the voltage pulse. Obviously, our experiment demonstrated that we can "turn off" the out-of-plane phonons of graphene at the nanoscale through voltage pulses applied to the STM tip. In Fig. 4e-4i, we show that it is also possible to "turn on" the out-of-plane phonons of graphene at the nanoscale through voltage pulses applied to the STM tip. In a nanoscale region, the "V-shape" spectrum of graphene monolayer can be reversibly changed to the gapped one, as shown in Fig. 4h. This result may arise from a competition between the tip-graphene interaction[37,38] and graphene-substrate interaction. During the process of applying the voltage pulses, such a competition, sometimes, could locally weaken the interaction between graphene and the substrate and, consequently, result in the emergence of the gapped tunneling spectra at the nanoscale region.

In summary, we systematically studied the two different tunneling spectra of the quasi-free-standing graphene monolayer and provided a consistent interpretation about the result. We demonstrated that the tunnelling spectra can be switched between the "V-shape" and the gapped features at the nanoscale, i.e., we can turn on and off the effect of the out-of-plane phonons of graphene, through voltage pulses applied to the STM tip. Moreover, our result indicated that graphite and graphene multilayer with a large stacking misorientation may be the best substrate to study the electronic properties of Dirac fermions of graphene systems in the STM studies.

**REFERENCES:**

1.  Miller, D. L., Kubista, K. D., Rutter, G. M., Ruan, M., de Heer, W. A., First, P. N., Stroscio, J. A. Observing the quantization of zero mass carriers in graphene. *Science* **324**, 924-927 (2009).

2.  Li, G., Luican, A., Andrei, E. Y. Scanning tunneling spectroscopy of graphene on graphite. *Phys. Rev. Lett.* **102**, 176804 (2009).

3.  Young, J. S., Alexander, F. O., Young, K., Yike, H., David, B. T., Phillip, N. F., Walt, A. H., Hongki, M., Shaffique, A., Mark, D. S., Allan, H. M. & Joseph, A. S. High-resolution tunnelling spectroscopy of a graphene quartet. *Nature* **467,** 185-189 (2010).

4.  Yin, L.-J. *et al*. Landau quantization of Dirac fermions in graphene and its multilayers. *Front. Phys.* **12**, 127208 (2017).

5.  Zhang, Y. *et al*. Giant phonon-induced conductance in scanning tunnelling spectroscopy of gate-tunable graphene. *Nature Phys.* **4**, 627 (2008).

6.  Zhang, Y., Brar, V. W., Girit, C., Zettl, A. & Crommie, M. F. Origin of spatial charge inhomogeneity in graphene. *Nature Phys.* **5**, 722 (2010).

7.  Kim, H. W., Ko, W., Ku, J., Jeon, I., Kim, D., Kwon, H., Oh, Y., Ryu, S., Kuk, Y., Hwang, S. W., Suh, H. Nanoscale control of phonon excitations in graphene. *Nature Commun.* **6**, 7528 (2015).

8.  Natterer, F. D., *et al*. Strong asymmetric charge carrier dependence in inelastic electron tunneling spectroscopy of graphene phonons. *Phys. Rev. Lett.* **114**, 245502 (2015).

9.  Levy, N., Burke, S. A., Meaker, K. L., Panlasigui, M., Zettl, A., Guinea, F., Castro Neto, A. H., Crommie, M. F. Strain-induced pseudo-magnetic fields greater than 300 Tesla in graphene nanobubbles. *Science* **329**, 544 (2010).

10. Yan, W., He, W.-Y., Chu, Z.-D., Liu, M., Meng, L., Dou, R.-F., Zhang, Y., Liu, Z., Nie, J.-C., He, L. Strain and curvature induced evolution of electronic band structures in twisted graphene bilayer. *Nature Commun.* **4**, 2159 (2013).

11. Zhao, L. *et al*. Visualizing Individual Nitrogen Dopants in Monolayer Graphene. *Science* **333**, 999 (2011).

12. Subramaniam, D. *et al*. Wave-function mapping of graphene quantum dots with soft confinement. *Phys. Rev. Lett.* **108**, 046801 (2012).

13. Hamalainen, S. K. *et al*. Quantum-confined electronic states in atomically well-defined graphene nanostructures. *Phys. Rev. Lett.* **107**, 236803 (2011).

14. Andrei, E. Y., Li, G. & Du, X. Electronic properties of graphene: a perspective from scanning tunneling microscopy and magnetotransport. *Rep. Prog. Phys.* **75,** 056501 (2012).

15. Wang, Y. *et al*. Observing atomic collapse resonances in artificial nuclei on graphene. *Science* **340**, 734 (2013).

16. Yin, L.-J., Jiang, H., Qiao, J.-B. & He, L. Direct imaging of topological edge states at a bilayer graphene domain wall. *Nature Commun.* **7**, 11760 (2016).

17. Calleja, F., Ochoa, H., Garnica, M., Barja, S., Navarro, J. J., Black, A., Otrokov, M. M., Chulkov, E. V., Arnau, A., Vazquez de Parga, A. L., Guinea, F., Miranda, R. Spatial variation of a giant spin-orbit effect induces electron confinement in graphene on Pb islands. *Nature Phys.* **11**, 43-47 (2015).

18. Li, G., Luican, A., Lopes dos Santos, J. M. B., Castro Neto, A. H., Reina, A., Kong, J. & Andrei, E. Y. Observation of Van Hove singularities in twisted graphene layers. *Nat. Phys.* **6,** 109-113 (2009).

19. Yan, W., Liu, M., Dou, R.-F., Meng, L., Feng, L., Chu, Z.-D., Zhang, Y., Liu, Z., Nie, J.-C., He, L. Angle-dependent van Hove singularities in a slightly twisted graphene bilayer. *Phys. Rev. Lett.* **109**, 126801 (2012).

20. Tapaszto, L., Dumitrica, T., Kim, S. J., Nemes-Incze, P., Hwang, C., Biro, L. P. Breakdown of continuum mechanics for nanometer-wavelength rippling of graphene. *Nature Phys.* **8**, 739-742 (2012).

21. Bai, K.-K., Zhou, Y., Zheng, H., Meng, L., Peng, H., Liu, Z., Nie, J.-C., He, L.

Creating one-dimensional nanoscale periodic ripples in a continuous mosaic graphene monolayer. *Phys. Rev. Lett.* **113**, 086102 (2014).

22. Zhao, Y. *et al.* Creating and probing electron whispering-gallery modes in graphene. *Science* **348**, 672 (2015).

23. Lee, J. *et al.* Imaging electrostatically confined Dirac fermions in graphene quantum dots. *Nature Phys.* **12**, 1032 (2016).

24. Gutierrez, C., Brown, L., Kim, C.-J., Park, J. & Pasupathy, A. N. Klein tunnelling and electron trapping in nanometre-scale graphene quantum dots. *Nature Phys.* **12**, 1069 (2016).

25. Bai, K. K., Qiao, J. B., Jiang, H., Liu, H. W. & He, L. Massless Dirac Fermions Trapping in a Quasi-one-dimensional npn Junction of a Continuous Graphene Monolayer. *Phys. Rev. B* **95**, 201406(R) (2017).

26. Ghahari, F., et al. An on/off Berry phase switch in circular graphene resonators. *Science* **356**, 845 (2017).

27. Jiang, Y., et al. Tuning a circular p-n junction in graphene from quantum confinement to optical guiding. arXiv: 1705.07346. To appear in Nature Nano.

28. Tao, C., Jiao, L., Yazyev, O. V., Chen, Y.-C., Feng, J., Zhang, X., Capaz, R. B., Tour, J. M., Zettl, A., Louie, S. G., Dai, H., & Crommie, M. F. Spatially resolving edge states of chiral graphene nanoribbons. *Nat. Phys.* **7,** 616-620 (2011).

29. Li, Y. Y., Chen, M. X., Weinert, M., & Li, L. Direct experimental determination of onset of electron-electron interactions in gap opening of zigzag graphene nanoribbons. *Nat. Commun.* **5,** 4311 (2014).

30. Magda, G. Z., Jin, X., Hagymasi, I., Vancso, P., Osvath, Z., Nemes-Incze, P., Hwang, C., Biro, L. P., & Tapaszto, L. Room temperature magnetic order on zigzag edges of narrow graphene nanoribbons. *Nature* **514,** 608-611 (2014).

31. Zhang, Y., Li, S. Y., Huang, H. Q., Li, W. T., Qiao, J. B., Wang, W. X., Yin, L. J.,

Bai, K. K., Duan, W. H., & He, L. Scanning Tunneling Microscopy of the π Magnetism of a Single Carbon Vacancy in Graphene. *Phys. Rev. Lett.* **117,** 166801 (2016).

32. Gonzalez-Herrero, H., Gomez-Rodriguez, J. M., Mallet, P., Moaied, M., Palacios, J. J., Salgado, C., Ugeda, M. M., Veuillen, J.-Y., Yndurain, F., & Brihuega, I. Atomic-scale control of graphene magnetism by using hydrogen atoms. *Science* **352,** 437-441 (2016).

33. Wehling, T. O., Grigorenko, I., Lichtenstein, A. I., Balatsky, A. V. Phonon-mediated tunneling into graphene. *Phys. Rev. Lett.* **101,** 216803 (2008).

34. Bai, K.-K., Wei, Y.-C., Qiao, J.-B., Li, S.-Y., Yin, L.-J., Yan, W., Nie, J.-C., He, L. Detecting giant electron-hole asymmetry in a graphene monolayer generated by strain and charged-defect scattering via Landau level spectroscopy. *Phys. Rev. B* **92,** 121405(R) (2015).

35. Li, S.-Y., Bai, K.-K., Yin, L.-J., Qiao, J.-B., Wang, W.-X., He, L. Observation of unconventional splitting of Landau levels in strained graphene. *Phys. Rev. B* **92,** 245302 (2015).

36. Khomyakov, P. A. *et al*. First-principle study of the interaction and charge transfer between graphene and metals. *Phys. Rev. B* **79**, 175425 (2009).

37. Klimov, N. N. *et al.* Electromechanical properties of graphene drumheads. *Science* **336**, 1557-1561 (2012).

38. Lim, H., Jung, J., Ruoff, R. S., Kim, Y. Structurally driven one-dimensional electron confinement in sub-5-nm graphene nanowrinkles. *Nature Commun.* **6**, 8610 (2015).

**Acknowledgments**

This work was supported by the National Natural Science Foundation of China (Grant Nos. 11674029, 11422430, 11374035), the National Basic Research Program of China (Grants Nos. 2014CB920903, 2013CBA01603), the program for New Century Excellent Talents in University of the Ministry of Education of China (Grant No. NCET-13-0054). L.H. also acknowledges support from the National Program for Support of Top-notch Young Professionals and support from "the Fundamental Research Funds for the Central Universities".

**Author contributions**

S.Y.L., K.K.B., W.J.Z., Y.W.L., W.X.W., Y.Z., J.B.Q., and L.J.Y. performed the STM experiments. S.Y.L. analyzed the data. L.H. conceived and provided advice on the experiment and analysis. L.H. and S.Y.L. wrote the paper. All authors participated in the data discussion.

**Competing financial interests:** The authors declare no competing financial interests.

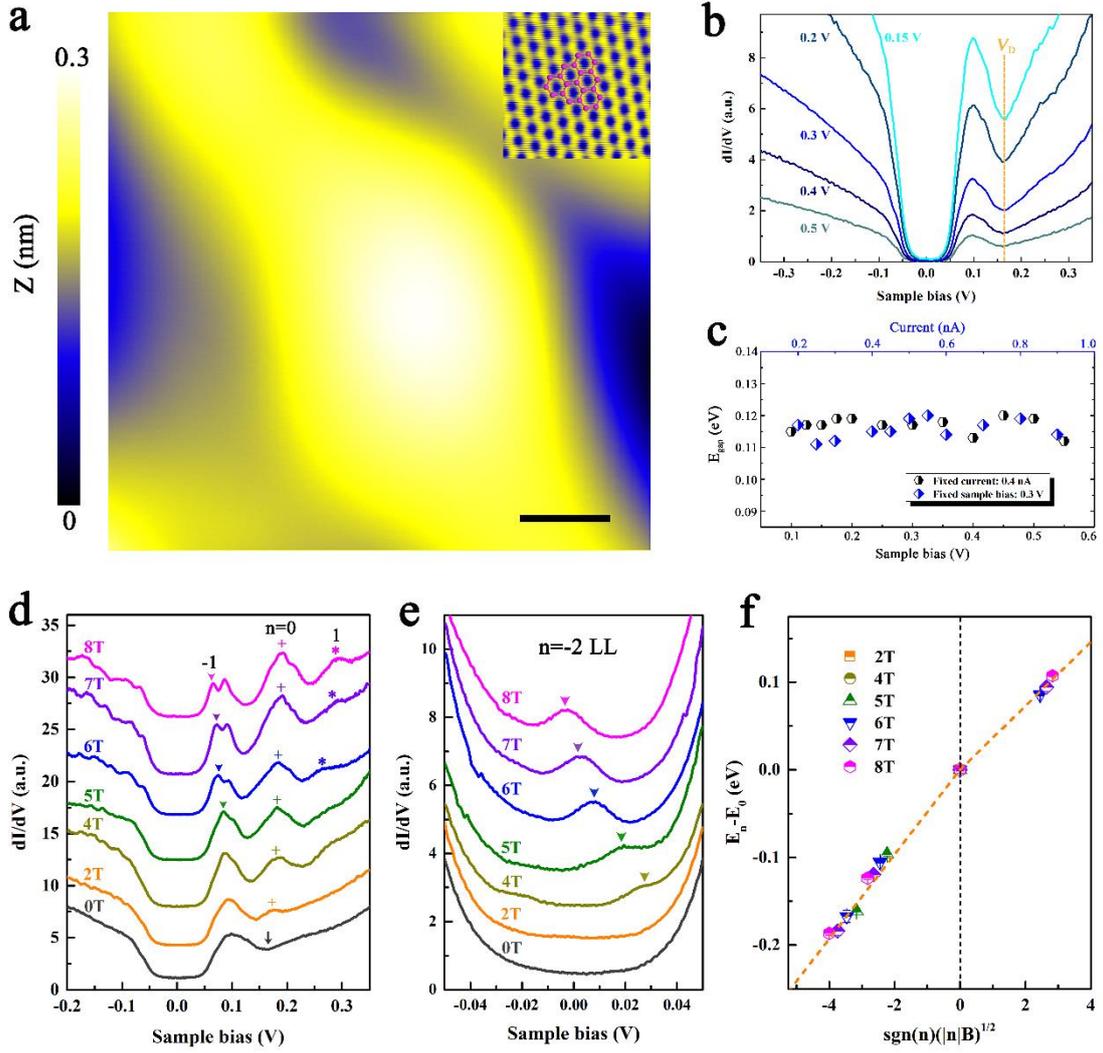

**Fig. 1. STM and STS of a suspended graphene region on a Rh foil. a.** A representative 30 nm ×30 nm STM image of a suspended graphene region on a Rh foil ($V_{sample}$ = 500 mV and $I$ = 0.4 nA). Scale bar: 5 nm. Inset: atomic-resolution STM image of this region showing hexagonal graphene lattices. **b.** The gapped tunneling spectra acquired in the suspended graphene monolayer region for different tip-sample heights (In the STM measurements, we use sample bias to tune the tip-sample height). The yellow dashed line marks the position of the Dirac point. **c.** Summarizing the gap values in the gapped tunneling spectra of the suspended graphene region for different experimental parameters. **d.** The tunneling spectra of the suspended graphene region recorded in different perpendicular magnetic fields. The Landau level peaks of n = 0, 1, -1 are labeled and the data are offset in the Y axis for clarity. The small gray arrow

marks the position of Dirac point in zero magnetic field. **e.** Enlarged tunneling conductance within the gap of the gapped tunneling spectra recorded under different magnetic fields. The Landau level n = -2 (marked by arrows) can be clearly seen inside the gap. **f.** The Landau level energies for different magnetic fields taken from panel d and panel e show a linear dependence against $\text{sgn}(n)(|n|B)^{1/2}$. The yellow dashed lines are linear fits of the data with Eq. (1), yielding the Fermi velocities $(1.33 \pm 0.03) \times 10^6$ m/s for electrons and $(1.00 \pm 0.03) \times 10^6$ m/s for holes respectively.

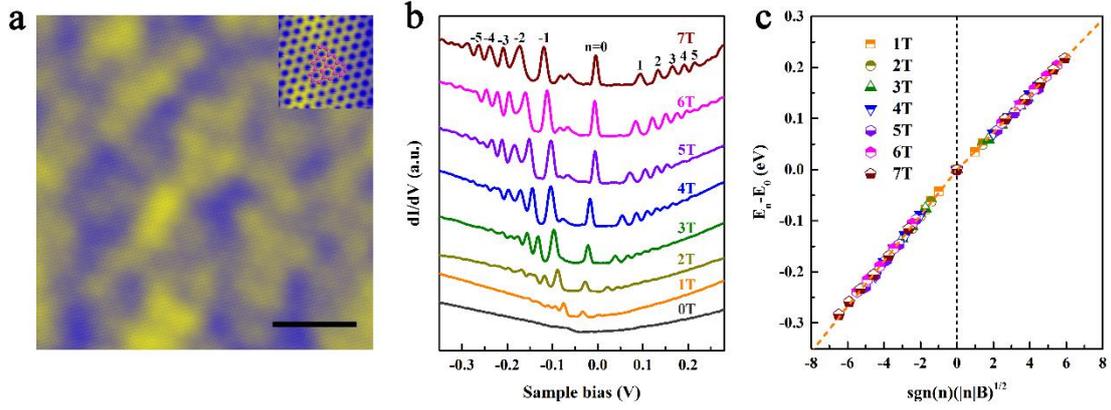

**Fig. 2. STM and STS of a decoupled graphene monolayer on multilayer graphene grown on a Rh foil. a.** A 20 nm ×20 nm STM image ($V_{sample}$ = 400 mV and $I$ = 0.4 nA) of the topmost graphene monolayer on multilayer graphene grown on Rh foil. Scale bar: 5 nm. The period of the moiré pattern generated between the topmost graphene monolayer and the underlying second layer is about 0.87 nm and the rotational angle between them is estimated to be about 16.3°. Inset: the atomic-resolution STM image of topmost graphene monolayer showing hexagonal graphene lattices. **b.** STS spectra taken from the decoupled graphene monolayer under different magnetic fields. The curves are offset on the Y axis for clarity and the LL indices of the massless Dirac fermions are labeled. **c.** The LL peak energies of graphene monolayer taken from panel b show a linear dependence against $sgn(n)(|n|B)^{1/2}$. The yellow dashed lines are the linear fits of LL peak energies with Eq. (1), yielding the Fermi velocities (1.23 ±0.01) ×$10^6$ m/s for electrons and (1.02 ±0.01) ×$10^6$ m/s for holes respectively.

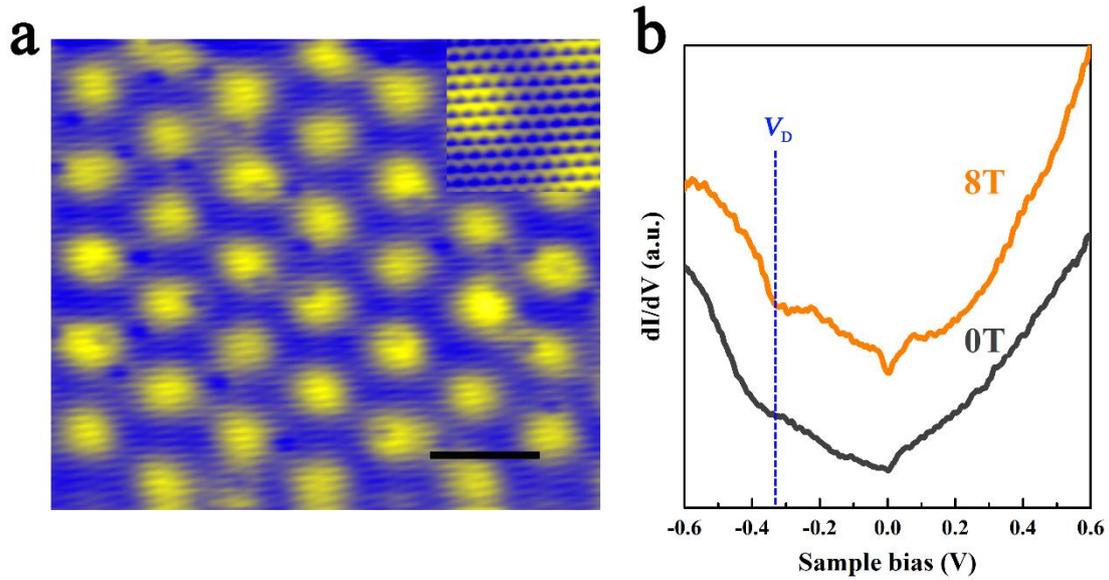

**Fig. 3. STM study of graphene monolayer on Cu(111) surface. a.** A 13 nm × 15 nm STM image ($V_{sample}$ = 610 mV and $I$ = 0.23 nA) of graphene monolayer on Cu(111) surface. Scale bar: 3 nm. There is a 5.3 ° twist angle between the graphene monolayer and the Cu(111) surface, resulting in the emergence of the moiré pattern with the period of 2.5 nm. Inset: the atomic-resolution STM image of the graphene monolayer. **b.** The STS spectra of the graphene monolayer on Cu(111) surface recorded in zero magnetic field and in the external magnetic field of 8 T. The blue dashed line marks the position of the Dirac point.

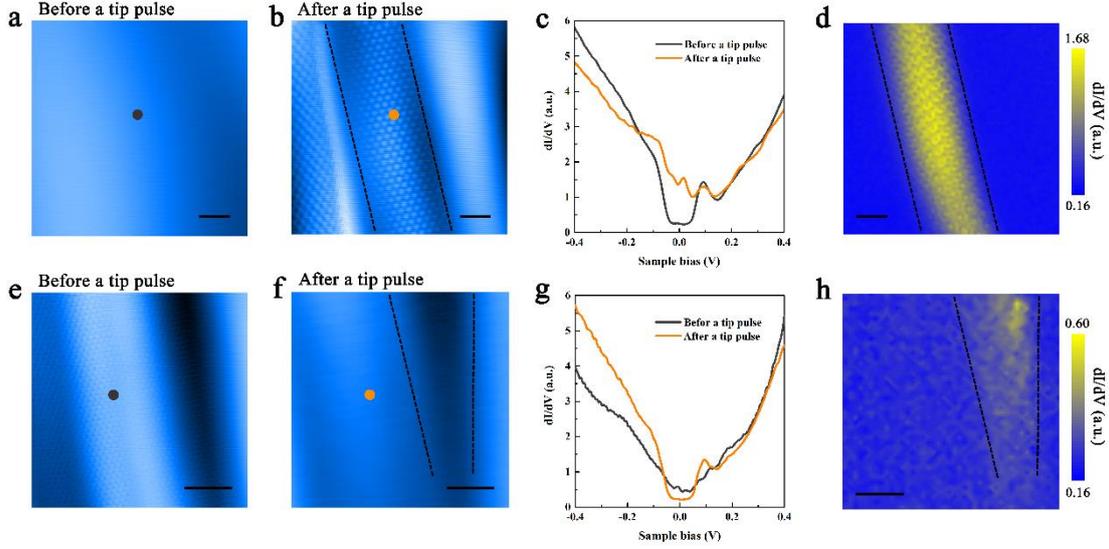

**Fig. 4. Turning on/off effects of the *K* and *K′* out-of-plane phonons in graphene. a.** A 7 nm ×7 nm STM image ($V_{sample}$ = 400 mV and $I$ = 0.4 nA) of a suspended graphene region on a Rh foil. Scale bar: 1 nm. **b.** The STM image of the same region as panel a after applying a 4 V tip pulse for 0.2 second. Scale bar: 1 nm. **c.** The gapped tunnelling spectrum is taken at the position marked by gray dot in panel a (before the tip pulse) and the other STS spectrum is recorded at the position marked by yellow dot in panel b after the tip pulse. **d.** dI/dV map obtained over the region shown in b at the bias voltage -9 mV. The yellow bright area between the two black dashed lines reflects the region that the effect of the *K* and *K′* out-of-plane phonons is suppressed after the tip pulse. **e, f.** 9 nm ×9 nm STM images ($V_{sample}$ = 400 mV and $I$ = 0.4 nA) obtained at the same graphene region before a tip pulse (e) and after a -4V tip pulse for 0.2 second (f), respectively. Scale bar: 2 nm. **g.** The grey solid line is the STS spectrum acquired at the position marked by gray dot in panel e (before the tip pulse) and the gapped tunneling spectrum is taken at the position marked by yellow dot in panel f after the tip pulse. **h.** dI/dV map obtained over the region shown in b at the bias voltage -8 mV. The effect of the *K* and *K′* out-of-plane phonons is turned on outside the yellow bright area between the two black dashed lines.